\documentclass[9pt,journal]{IEEEtran}
\ifCLASSINFOpdf
\else
\fi
\newtheorem{theorem}{Theorem}
\newtheorem{lemma}{Lemma}
\newtheorem{proposition}{Proposition}

\newtheorem{example}{Example}
\newtheorem{remark}{Remark}

\begin{document}
%
\title{Optimal dynamic allocation of collaborative servers in two station tandem systems}
%
%
%
\author{Ioannis~Papachristos
        and~Dimitrios~G.~Pandelis
\thanks{This research was supported by the ECSEL Joint Undertaking under grant
agreement no. 737459. This Joint Undertaking receives support from the
European Union's Horizon 2020 Research and Innovation Program and Germany,
Austria, France, Czech Republic, The Netherlands, Belgium, Spain,Greece,
Sweden, Italy, Ireland, Poland,Hungary, Portugal, Denmark, Finland,
Luxembourg, Norway, and Turkey.}
\thanks{The authors are with the Department
of Mechanical Engineering, University of Thessaly, 38334 Volos, Greece (e-mail: iopapahr@uth.gr; d\_ pandelis@mie.uth.gr).}}
%
%

\markboth{t\MakeLowercase{his article has been accepted for publication in} IEEE Transactions on Automatic Control. C\MakeLowercase{itation information: DOI 10.1109/TAC.2018.2852604.}}
{Shell \MakeLowercase{\textit{et al.}}: Bare Demo of IEEEtran.cls for Journals}
%


\IEEEpubid{0018-9286 (c) 2018 IEEE. Personal use is permitted, but republication/redistribution requires IEEE permission.}


\maketitle


\begin{abstract}
We consider two-stage tandem queueing systems with one dedicated server in each station and a flexible server that can serve both stations. We assume exponential service times, linear holding costs accrued by jobs present in the
system, and a collaborative work discipline. We seek optimal server allocation strategies for systems without external arrivals (clearing systems). When the combined rate of collaborating servers is less than the sum of their individual rates (partial collaboration), we identify conditions under which the optimal server allocation strategy is non-idling and has a threshold-type structure. Our results extend previous work on systems with additive service rates. When the aforementioned conditions are not satisfied we show by examples that the optimal policy may have counterintuitive
properties, which is not the case when a fully collaborative service discipline is assumed. We also obtain novel results for any type of collaboration when idling policies may be optimal.
\end{abstract}

%
\IEEEpeerreviewmaketitle

\section{Introduction}
%
%
%
%
\IEEEPARstart{W}{e} study two-station tandem queueing systems with one dedicated server in each station and one flexible server that is trained to work in both stations. Our objective is to determine properties of server allocation strategies that minimize expected linear holding costs for Markovian systems without arrivals (clearing systems). The problem we consider is motivated by the use of cross-trained workers in manufacturing systems in order to cope with variability in demand, processing times, and operating conditions. Unlike traditional settings where each worker could perform a single task, cross-trained workers can be assigned to tasks where they are needed the most resulting in increased efficiency in the form of higher throughput, lower inventory, etc.  Hopp and Van Oyen \cite{Hopp1} have provided  a literature survey on workforce flexibility as well as a framework for evaluating a flexible workforce in an organization. A more recent survey can be found in Andradottir, Ayhan, and Down \cite{Andr1} along with design guidelines for eliminating bottlenecks.

The search for server allocation policies that minimize holding costs has focused on two-stage systems and exponential service times. Ahn, Duenyas, and Zhang \cite{Ahn1} and  Ahn, Duenyas, and Lewis \cite{Ahn2} identified conditions for the optimality of exhaustive policies in clearing systems and systems with arrivals, respectively, with two flexible servers. Extensions for \cite{Ahn1},\cite{Ahn2} were obtained by Schiefermayr and Weichbold \cite{Schiefer} and Weichbold and Schiefermayr \cite{Weichbold}. Kirkizlar, Andradottir, and Ayhan \cite{Kirkiz2} also studied a model with two flexible servers where, in addition to holding costs, a profit is earned for each job completion. They showed that the optimal strategy is characterized by a threshold and determined the value of this threshold.

Our work is mostly related to models with dedicated servers as well.
Farrar \cite{Farrar1},\cite{Farrar2} considered two versions of a clearing system with dedicated servers in each station and one extra server. In the constrained version the extra server can only work in the upstream station, whereas in the unconstrained version the server can work in both stations. He showed that for both versions the optimal policy is characterized by a switching curve with slope greater than or equal to -1, which implies that if the flexible server is idled or assigned to the downstream station, its allocation does not change if a job joins the queue from upstream ({\it transition monotone} policy). Pandelis \cite{PandelisMMOR} extended the results in \cite{Farrar1},\cite{Farrar2} to the case when jobs may leave the system after completing service in the first station. The same structure of the optimal policy was obtained by Wu, Lewis, and Veatch \cite{Wu1} where it was assumed that the servers have varying speeds and the processing requirements are the same in both stations.  Wu, Down, and Lewis \cite{Wu2} showed the optimality of a policy with similar monotonicity properties for the previous model with arrivals and no dedicated server in the upstream station and Pandelis \cite{PandelisORL} extended this result to the case when jobs may not require service in the downstream station and processing requirements are not the same in each station. Finally, Pandelis \cite{PandelisPEIS} studied a model with server operating costs in addition to holding costs and identified conditions under which the switching-curve structure of the optimal policy is preserved. With the exception of \cite{PandelisMMOR} (constrained version), a common assumption in all of the papers cited in this paragraph was that different servers could collaborate to work on the same job, in which case the total service rate was equal to the sum of their individual rates ({\it fully collaborative} servers). Moreover, a non-idling discipline for at least the dedicated servers was assumed. Both of these conditions were relaxed by Pandelis \cite{PandelisNRL}, who provided conditions under which non-idling policies that are characterized by a single switching curve are optimal for clearing systems under a non-collaborative service discipline.

\IEEEpubidadjcol
In this paper we derive properties of optimal policies under the assumption that the combined service rate of two servers collaborating to work on the same job is less than the sum of their individual service rates ({\it partially collaborative} servers). Some of our results are novel for any type of collaboration (partial and full), while the rest extend previous work on fully collaborative servers. Situations with partial collaboration arise when for some reason (e.g., servers sharing resources when collaborating) it is not possible for each server to achieve full performance. The assumption of non-additive service rates has also been used in the work of Ahn and Lewis \cite{Ahn3} who studied the problem of optimal routing and flexible server allocation to two parallel queues. In addition to partially collaborating servers (subadditive rates) they considered the case when collaboration increases the servers' efficiency, that is, their combined service rate is larger than the sum of their individual rates (superadditive rates). Models with non-additive rates for tandem systems with throughput maximization as the objective were studied by Andradottir, Ayhan, and Down \cite{Andr6} (subadditive rates), Andradottir, Ayhan, and Down \cite{Andr5} and Wang, Andradottir, and Ayhan \cite{Wang} (superadditive rates). For our model it will become evident from the ensuing analysis that the problem with  superadditive service rates is equivalent to a problem with fully collaborative servers, so we do not consider this case.

The problem is formulated in Section II and analyzed in Section III. When non-idling policies are optimal we extend results from past literature by providing conditions on service rates under which the structure of the optimal policy for fully collaborative servers is preserved under partial collaboration. When these conditions are not satisfied we show by examples that the optimal server allocation may not possess the same structure and in fact be quite counterintuitive. When idling policies are optimal we obtain properties of the optimal policy that are novel for any type of collaboration. Specifically, we provide an asymptotic characterization of the optimal policy for a large number of jobs in the downstream station, and in case of no dedicated server in the downstream station we show that the optimal allocation is determined by  a strict priority rule for the flexible server and a monotone switching curve for the dedicated server. We conclude in Section IV. Most proofs are given in an Appendix; only short proofs of basic results are included in the main text.

\section{Problem Formulation and Preliminaries}
\IEEEpubidadjcol
We study two-stage tandem queueing systems with a number of jobs initially present and no further arrivals. After being served in the upstream station (Station 1), jobs move to the downstream station (Station 2) where they receive additional service and then leave the system. Each job in Station $i$, $i=1,2$, waiting or in service, incurs linear holding costs at a strictly positive rate $h_i$. There are two dedicated servers, one for each station, that are trained to work only in their corresponding station, and one flexible server that can attend both stations. We allow preemptions at times of service completions and assume that there is no cost or delay when the flexible server moves from station to station. We assume exponential service times with rates $\nu_1,\nu_2$ for jobs processed by the dedicated server and $\mu_1,\mu_2$ for jobs processed by the flexible server in Station 1 and 2, respectively. Two servers can work simultaneously on different jobs in the same station, as well as collaborate to work on the same job. When the collaboration takes place in Station $i$, $i=1,2$, the service rate is equal to $\nu_i+\xi_i$, where $\nu_i+\xi_i>\mu_i$ and $0<\xi_i\leq\mu_i$, with equality corresponding to full collaboration. Our objective is to find a server allocation strategy that minimizes the total expected holding cost until the system is cleared of all jobs.

We formulate the problem as a Markov decision process with state space $\{(x_1,x_2):~x_1,x_2\geq0\}$, where $x_i$, $i=1,2$, is the number of jobs in Station $i$, including those in service. Starting from state $(x_1,x_2)$, we denote by $V(x_1,x_2)$ the minimum total expected holding cost until the system empties, with $V(0,0)=0$. Instead of the continuous time problem, we study an equivalent discrete time problem obtained by uniformization (see, e.g., \cite{Serfozo}), where without loss of generality we assume $\nu_1+\nu_2+\mu_1+\mu_2+\xi_1+\xi_2=1$. Then, with ${\cal A}(x_1,x_2)$ denoting the set of feasible service rates in state $(x_1,x_2)$, we get the following optimality equation.
\begin{eqnarray}
V(x_1,x_2)&=&h_1x_1+h_2x_2\nonumber\\
&&\hspace{-1cm}+\min_{(\rho_1,\rho_2)\in {\cal A}(x_1,x_2)}W_{\rho_1,\rho_2}(x_1,x_2),
\end{eqnarray}
where
\begin{eqnarray}
W_{\rho_1,\rho_2}(x_1,x_2)&=&\rho_1V(x_1-1,x_2+1)\nonumber\\
&&\hspace{-2cm}+\rho_2V(x_1,x_2-1)+(1-\rho_1-\rho_2)V(x_1,x_2).
\end{eqnarray}
Note that if $x_1=0$ (resp. $x_2=0$), we get $V(-1,x_2+1)$ (resp. $V(x_1,-1)$) in (2), which are terms that have not been formally defined. However, this is not a problem because the only feasible rate is $\rho_1=0$ (resp. $\rho_2=0$).

Before proceeding to the characterization of the optimal policy, we give preliminary results that will be used in the proof of the main results of this section. Lemma 1 states that the minimum expected cost increases with the number of jobs in the system. It can be proved by induction on the number of jobs (as in \cite{PandelisNRL}) or by straightforward sample path arguments.
\begin{lemma}
$V(x_1,x_2)$ is increasing in $x_1$ and $x_2$.
\end{lemma}
With $A^+=\max(0,A)$, $A^-=\min(0,A)$ denoting the positive and negative part of $A$, Lemma 2 gives an auxiliary result that will be used in comparisons that determine the optimal server allocation. Its proof can be obtained easily by contradiction.
\begin{lemma}
Suppose that $A-B=G+\alpha (A-B)+\beta(A^--B^-)+\gamma(A^+-B^+)$, where $A,B,G,\alpha,\beta,\gamma$ are real numbers with $\alpha+\beta<1$ and $\alpha+\gamma<1$. Then, $A-B$ and $G$ have the same sign.
\end{lemma}

\section{The Optimal Policy}
When one of the two queues is empty of jobs, it is clear that the optimal policy allocates the maximum possible service rate to the nonempty queue. When there is one job, the dedicated and the flexible server work together on that job, otherwise they work on separate jobs. Therefore,
\begin{eqnarray}
V(1,0)&=&\frac{h_1}{\nu_1+\xi_1}+V(0,1),\\
V(x_1,0)&=&\frac{h_1x_1}{\nu_1+\mu_1}+V(x_1-1,1),~~x_1>1\\
V(0,1)&=&\frac{h_2}{\nu_2+\xi_2}+V(0,0),\\
V(0,x_2)&=&\frac{h_2x_2}{\nu_2+\mu_2}+V(0,x_2-1),~~x_2>1.
\end{eqnarray}
When there are jobs in station $i$, $i=1,2$, assuming an initial allocation $\rho_1,\rho_2$ the incentive to allocate additional rate $\rho$ to that station is $\rho f(x_1,x_2)$ for $i=1$ and $-\rho g(x_1,x_2)$ for $i=2$, where
\begin{displaymath}
f(x_1,x_2)=V(x_1,x_2)-V(x_1-1,x_2+1),~~x_1\geq 1,~x_2\geq 0,
\end{displaymath}
\begin{displaymath}
g(x_1,x_2)=V(x_1,x_2-1)-V(x_1,x_2),~~x_1\geq 0,~x_2\geq 1,
\end{displaymath}
and $g(x_1,x_2)<0$ by Lemma 1. Therefore, the optimal policy should allocate as much service rate as possible to Station 2 and  depending on the sign of $f(x_1,x_2)$, either allocate as many resources as possible to Station 1 or no resources at all. These properties of the optimal policy are formally  described in the following propositions.
\begin{proposition}
For given $\rho_1$, $W_{\rho_1,\rho_2}(x_1,x_2)$ is minimized by maximizing $\rho_2$.
\end{proposition}
\begin{proposition}
For given $\rho_2$, $W_{\rho_1,\rho_2}(x_1,x_2)$ is minimized by maximizing $\rho_1$ if $f(x_1,x_2)\geq 0$ and by $\rho_1=0$ if  $f(x_1,x_2)<0$.
\end{proposition}
Propositions 1 and 2 imply that the optimal policy does not idle any server when $f(x_1,x_2)\geq 0$, whereas in the opposite case it idles the dedicated server of Station 1 and assigns the flexible server to Station 2 to work along with its dedicated server.
\begin{remark}
Propositions 1 and 2 hold for any form of collaboration. Then, assuming superadditive service rates, that is, $\xi_i>\mu_i$, $i=1,2$, the optimal policy would always have the flexible server collaborating with one of the dedicated servers, say server $i$, to work on the same job, resulting in a total service rate of $\nu_i+\xi_i$. This is equivalent to an additive service rate model with rates $\xi_1,\xi_2$ for jobs served by the flexible server in Station 1 and 2, respectively.
\end{remark}
\subsection{Optimality of non-idling policies}
In this section we investigate the structure of optimal policies for $h_1\geq h_2$, which is a necessary and sufficient condition for the optimality of non-idling policies. Intuitively, when it is not cheaper to have jobs in Station 1 compared to Station 2, it is reasonable not to idle resources to keep jobs upstream. The necessity of the condition is proved in the next section (Theorem 4). The sufficiency is a consequence of the following lemma, whose proof can be obtained by induction on the number of jobs in Station 2 (see \cite{PandelisNRL}) or by sample path arguments.
\begin{lemma}
Let $h_1\geq h_2$. Then, $f(x_1,x_2)>0$ for all $x_1\geq 1$, $x_2\geq 0$.
\end{lemma}

Having established the optimality of non-idling policies we give conditions under which properties of the optimal policy that have been shown to hold for models with fully collaborative servers also hold with partial collaboration. Specifically, we show that the optimal allocation of the flexible server is determined by a switching curve. Moreover, for systems with one dedicated server we show that under certain conditions strict priority rules are optimal. For fully collaborative servers these properties had been proved in \cite{PandelisPEIS} for a more general model that included server operating costs. The switching curve property had also been proved earlier for models with $\mu_1=\mu_2$ (\cite{Farrar2},\cite{PandelisMMOR},\cite{Wu1}).

Because idling a server cannot be optimal, the decision to be made is where to assign the flexible server. Propositions 1 and 2 also imply that when there are at least two jobs in the station to which the flexible server is assigned, the two servers should work on separate jobs rather than collaborate on the same job.  Therefore, taking also into account that $\nu_1+\nu_2+\mu_1+\mu_2+\xi_1+\xi_2=1$, we get from (1) and (2) the following optimality equations.
\begin{eqnarray}
V(1,1)&=&h_1+h_2+\nu_1V(0,2)+\nu_2V(1,0)\nonumber\\
&&\hspace{-1cm}+(\mu_1+\mu_2)V(1,1)+\min\{\xi_1V(0,2)+\xi_2V(1,1),\nonumber\\
&&\hspace{-1cm}\xi_2V(1,0)+\xi_1V(1,1)\},
\end{eqnarray}
and for $x_1,x_2>1$,
\begin{eqnarray}
V(1,x_2)&=&h_1+h_2x_2+\nu_1V(0,x_2+1)\nonumber\\
&&\hspace{-1cm}+\nu_2V(1,x_2-1)+(\xi_2+\mu_1)V(1,x_2)\nonumber\\
&&\hspace{-1cm}+\min\{\xi_1V(0,x_2+1)+\mu_2V(1,x_2),\nonumber\\
&&\hspace{-1cm}\mu_2V(1,x_2-1)+\xi_1V(1,x_2)\},\\
V(x_1,1)&=&h_1x_1+h_2+\nu_1V(x_1-1,2)\nonumber\\
&&\hspace{-1cm}+\nu_2V(x_1,0)+(\xi_1+\mu_2)V(x_1,1)\nonumber\\
&&\hspace{-1cm}+\min\{\mu_1V(x_1-1,2)+\xi_2V(x_1,1),\nonumber\\
&&\hspace{-1cm}\xi_2V(x_1,0)+\mu_1V(x_1,1)\},\\
V(x_1,x_2)&=&h_1x_1+h_2x_2+\nu_1V(x_1-1,x_2+1)\nonumber\\
&&\hspace{-1cm}+\nu_2V(x_1,x_2-1)+(\xi_1+\xi_2)V(x_1,x_2)\nonumber\\
&&\hspace{-1cm}+\min\{\mu_1V(x_1-1,x_2+1)+\mu_2V(x_1,x_2),\nonumber\\
&&\hspace{-1cm}\mu_2V(x_1,x_2-1)+\mu_1V(x_1,x_2)\},
\end{eqnarray}
where the first and second terms in braces correspond to the assignment of the flexible server to the first and second station, respectively. Next, we define a set of functions that determine the optimal decision in each state. For $x_1\geq1$, $x_2\geq0$,
\begin{equation}
d(x_1,x_2)=\mu_1f(x_1,x_2)+\mu_2g(x_1,x_2),
\end{equation}
\begin{equation}
\tilde d(1,x_2)=\xi_1f(1,x_2)+\mu_2g(1,x_2),
\end{equation}
\begin{equation}
\hat d(x_1,1)=\mu_1f(x_1,1)+\xi_2g(x_1,1),
\end{equation}
\begin{equation}
\bar d(1,1)=\xi_1f(1,1)+\xi_2g(1,1),
\end{equation}
with $g(x_1,0)=0$. Function $d(x_1,x_2)$ is derived by subtracting the first from the second term in curly brackets in (10). Therefore, its sign determines the optimal allocation for the flexible server when there are at least two jobs in each station: assign the server upstream if $d(x_1,x_2)\geq 0$, and downstream otherwise. Similarly, $\hat d(x_1,1)$ is the decision function when there is one job in the downstream station and at least two jobs upstream, $\tilde d(1,x_2)$ is the decision function when there is one job in the upstream station and at least two jobs downstream, and $\bar d(1,1)$ is the decision function when there is one job in each station.

The main result of this section is given in Theorem 1. Its proof requires the use of the three lemmas that follow. Lemmas 4 and 5 are used to prove that the optimal policy is determined by a single switching curve, formally defined in the statement of the theorem, and Lemma 6 to obtain a lower bound on its slope.
\begin{lemma}
Let $\nu_2\geq\mu_2$. Then\\
i) For $x_2\geq 0$,
$\tilde d(1,x_2+1)<\tilde d(1,x_2)$.\\
ii) $\lim_{x_2\rightarrow\infty}\tilde d(1,x_2)=-\infty$.
\end{lemma}
\begin{lemma}
Let $\nu_2\geq\mu_2$ and $\mu_1\geq \mu_2$. Then\\
i) For $x_1\geq 1$, $x_2\geq 0$,
$d(x_1,x_2+1)<d(x_1,x_2)$.\\
ii) For $x_1\geq 1$,
$\lim_{x_2\rightarrow\infty}d(x_1,x_2)=-\infty$.
\end{lemma}
\begin{lemma}
Let $\nu_2\geq\mu_2$ and $\mu_1\geq \mu_2$. Then\\
i) For $x_2\geq 1$,
$d(2,x_2)\geq0\Longrightarrow d(2,x_2)\geq d(1,x_2+1)$.\\
ii) For $x_2\geq 2$,
$\tilde d(1,x_2)\geq0\Longrightarrow\tilde d(1,x_2)\leq d(2,x_2-1)$.\\
iii) For $x_1\geq2$, $x_2\geq 2$,
$d(x_1,x_2)\geq0\Longrightarrow d(x_1,x_2)\leq d(x_1+1,x_2-1)$.
\end{lemma}

The proofs of the lemmas, which can be found in the appendix, are based on induction arguments applied to recursive equations for decision functions $\tilde d(1,x_2)$, $\hat d(x_1,1)$, and $d(x_1,x_2)$. It can be seen that the proofs require a lot of complicated, nontrivial technical arguments, especially for comparing the values of decision functions for small $x_1,x_2$. This is due to the fact that the recursive expression for each decision function involves other decision functions as well. This is not the case for fully collaborative servers where $d(x_1,x_2)$ is the decision function for all $x_1,x_2\geq1$, so we have to deal with a single recursive equation involving only one decision function. As a result the proofs are short and straightforward (see \cite {PandelisPEIS} where a similar equation is derived for a model with operating costs). Note that compared to fully collaborative servers the assumption of partial collaboration only matters for states with one job in one or both stations, because otherwise the optimal policy assigns the servers to different jobs, which is equivalent to full collaboration. Therefore, we see that this minor difference between the two models results in a greatly disproportionate increase in the complexity of the technical analysis for partially collaborative servers.
\begin{theorem}
Assume $h_1\geq h_2$, $\nu_2\geq\mu_2$, $\mu_1\geq \mu_2$, and partially collaborative servers. Then, for each $x_1\geq 1$, there exists an integer
$t(x_1)\geq 1$ such that the optimal policy assigns the flexible server to Station 2 (resp. 1) when $x_2\geq t(x_1)$ (resp. $x_2<t(x_1)$). Moreover, the slope of $t(x_1)$ is at least -1.
\end{theorem}
\begin{IEEEproof}
To prove the existence part, we first consider $x_1=1$. If $\bar d(1,1)<0$, then for $x_2\geq 2$ we have $\tilde d(1,x_2)<0$ because of $\tilde d(1,1)<\bar d(1,1)$ and Lemma 4(i). Therefore, the optimal policy assigns the flexible server to the downstream station for any number of jobs in that station, that is, $t(1)=1$. Otherwise, let $m=\min\{x_2\geq 2:~\tilde d(1,x_2)<0\}$, noting that the existence of this minimum is guaranteed by Lemma 4(ii). Then, $t(1)=m$ because Lemma 4(i) implies $\tilde d(1,x_2)<0$ for $x_2\geq m$. For $x_1>1$ the statement of the theorem is proved similarly by using the fact that $d(x_1,1)<\hat d(x_1,1)$ and Lemma 5.

The fact that the slope of $t(x_1)$ is at least -1 is a consequence of parts (ii) and (iii) of Lemma 6, from which it follows that if the decision function is negative for some state $(x_1,x_2)$, it is also negative for $(x_1-1,x_2+1)$.
\end{IEEEproof}

As seen form its statement, we were able to prove Theorem 1 under conditions $\nu_2\geq\mu_2$ and $\mu_1\geq\mu_2$. The first one implies that the specialist (dedicated server) in Station 2 is not slower than the generalist (flexible) server, which is a reasonable assumption. However, this is not the case with the second condition which seems arbitrary. An interesting question is whether the two conditions are crucial for the validity of the results, or they were just needed for the arguments of the proofs to work. To answer this question we obtained numerical results that illustrate the structure of the optimal policy when either one or both of the conditions are violated. For each of the three cases we created 100,000 problem instances with randomly generated values for service and holding cost rates and computed the optimal server allocation for each one. When only one of the conditions was violated, all of our results were in agreement with Theorem 1. Moreover, we observed that the switching curve was nondecreasing in all instances. When both conditions were violated, the optimal policy was still determined by a unique switching curve, but we found instances with switching curves having a portion with slope less than -1. One such instance is given in the following example.
\begin{example}
Let $\nu_1=0.8$, $\mu_1=0.6$, $\xi_1=0.03$, $\nu_2=0.6$, $\mu_2=8$, $\xi_2=7.43$, $h_1=16$, and $h_2=1.5$. When there are three jobs in each station, the optimal policy assigns the flexible server to Station 2. However, if a job completes its service in Station 1 and joins Station 2, then, contrary to intuition, the flexible server is transferred to Station 1.
\end{example}

In all of the instances with a counterintuitive optimal policy the service rate of the flexible server in Station 2 is much larger than any other service rate. Therefore, when the flexible server is assigned to Station 2, there is a very high probability that the first event to occur is a service completion there. Otherwise, a completion in Station 1 is the most probable event, but not with such a high probability.  Regarding the optimal policy, it is affected by the service discipline when there are a few jobs in the system. For fully collaborative servers we observed that it favors emptying Station 2 first. On the other hand, partial collaboration seems to change significantly the dynamics of the system and emptying Station 1 first is favored. As is the case with Example 1, the peculiar behavior of the optimal policy for partial collaboration occurs when a service completion in Station 1, which is a very low probability event, moves the state of the system closer to the target and the flexible server is moved to Station 1 in further support of reaching this target. This is not the case with full collaboration when the low probability event moves the state away from the target, so the flexible server remains in Station 2.

Further characterizations of the optimal policy can be obtained for systems with no dedicated server in one of the stations. For such systems, depending on the expected holding cost savings resulting from the assignment of the flexible server to a specific station, the optimal allocation of the flexible server may be explicitly determined by a strict priority rule. The aforementioned savings are equal to $\mu_1(h_1-h_2)$ for Station 1 (because the completed job leaves Station 1 and joins Station 2) and $\mu_2h_2$ for Station 2 (because the completed job leaves Station 2 and the system). Then, it is reasonable to expect that if the station with no dedicated server is also the one with the largest expected savings, the optimal policy would give priority to that station. We show later by example that this intuitive priority rule, which is optimal in the full collaboration case, may not be optimal for partially collaborative servers.

The following theorem gives properties of the optimal policy when there is no dedicated server assigned to Station 1. Note that in this case non-idling policies are optimal for any values of holding cost rates so condition $h_1\geq h_2$ is not needed.
\begin{theorem}
Assume $\nu_1=0$ and partially collaborative servers in Station 2. Then\\
i) When $\mu_1(h_1-h_2)<\mu_2h_2$ and $\nu_2\geq\mu_2$, for each $x_1\geq 1$, there exists an integer
$t(x_1)\geq 1$ such that the optimal policy assigns the flexible server to Station 2 (resp. 1) when $x_2\geq t(x_1)$ (resp.
$x_2<t(x_1)$). Moreover, the slope of $t(x_1)$ is at least -1.\\
ii) When $\mu_1(h_1-h_2)\geq\mu_2h_2$, the optimal policy assigns the flexible server to Station 1 for all $x_1\geq1$.
\end{theorem}
\begin{IEEEproof}
When there is no dedicated server assigned to Station 1, the optimality equations are given by (9) and (10) for all $x_1\geq1$ and $x_2=1$, $x_2>1$, respectively, with $\nu_1=\xi_1=0$. Therefore, the decision function whose sign determines the optimal policy is
$\hat d(x_1,1)$ for one job downstream and $d(x_1,x_2)$ otherwise.
We prove that the optimal policy is characterized by a switching curve $t(x_1)$ by showing that $d(x_1,x_2)$, $x_1\geq 1$, is decreasing in $x_2$ and its limit as $x_2\rightarrow\infty$ is $-\infty$. For $t(x_1)$ having a slope equal to at least -1 we prove part (iii) of Lemma 6 for $x_1\geq1$. For part (ii) we show by induction on $x_1$ that $\lim_{x_2\rightarrow\infty}d(x_1,x_2)\geq0$ when $\mu_1(h_1-h_2)\geq\mu_2h_2$. Because of lack of space we omit the details of the proof and refer interested readers to \cite{Papachristos}.
\end{IEEEproof}

As with Theorem 1, we conducted a numerical investigation to see whether condition $\nu_2\geq\mu_2$ is needed for the validity of part (i) of Theorem 2 by examining 100,000 test cases with $\nu_2<\mu_2$. We found the structure of the optimal policy for all of them to be in agreement with the theorem. In addition, $t(x_1)$ was nondecreasing for all cases.

When there is no dedicated server assigned to Station 2, the optimal policy is characterized in the following theorem.
\begin{theorem}
Assume $h_1\geq h_2$, $\nu_2=0$, $\mu_1\geq\mu_2$, and partially collaborative servers in Station 1. Then\\
i) For each $x_1\geq 1$, there exists an integer
$t(x_1)\geq 1$ such that the optimal policy assigns the flexible server to Station 2 (resp. 1) when $x_2\geq t(x_1)$ (resp.
$x_2<t(x_1)$). Moreover, the slope of $t(x_1)$ is at least -1.\\
ii) When $\mu_1(h_1-h_2)\leq\mu_2h_2$, the optimal policy assigns the flexible server to Station 2, that is, $t(x_1)=1$.
\end{theorem}
\begin{IEEEproof}
The optimality equations in this case are given by (8) and (10) for all $x_2\geq1$ and $x_1=1$, $x_1>1$, respectively, with $\nu_2=\xi_2=0$. Therefore, the decision function is $\tilde d(1,x_2)$ for one job in the first station, and $d(x_1,x_2)$ otherwise.
To prove the existence of $t(x_1)$ we show that $\tilde d(1,x_2)$ and $d(x_1,x_2)$ are decreasing and their limit as $x_2\rightarrow\infty$ is $-\infty$. For $t(x_1)$ having a slope equal to at least -1 we prove the properties cited in the statement of Lemma 6. Finally, to prove part (ii) we show that $\tilde d(1,1)<0$ and $d(x_1,1)<0$ when $\mu_1(h_1-h_2)\leq\mu_2h_2$. All the details can be found in \cite{Papachristos}.
\end{IEEEproof}

Similarly to Theorem 1, we used numerical experiments to examine the effect of condition $\mu_1\geq\mu_2$ on the validity of Theorem 3. When the condition is violated, we found that part (i) holds for $\mu_1(h_1-h_2)>\mu_2 h_2$. On the other hand, the priority rule in part (ii) may not be optimal when $\mu_1<\mu_2$. This is illustrated in the following example.
\begin{example}
Let $\nu_1=1.3$, $\mu_1=0.9$, $\xi_1=0.1$, $\mu_2=7.7$, $h_1=11.4$, and $h_2=1.2$, so that $\mu_1(h_1-h_2)<\mu_2 h_2$. However, there are states (e.g., $x_1=3$, $x_2=1$) for which the flexible server is assigned to Station 1.
\end{example}

\subsection{Optimality of idling policies}
In this section we consider the case with larger holding cost rate in Station 2, that is, $h_1< h_2$. We show that idling policies are optimal when there is a sufficiently large number of jobs downstream. This is reasonable because it may be better to prevent jobs from joining the more expensive Station 2 by not assigning any resources to Station 1. In related previous work with fully collaborative servers (\cite{Farrar2},\cite{PandelisMMOR},\cite{Wu1},\cite{PandelisPEIS}) the case $h_1< h_2$ was not considered separately because the search for an optimal policy was restricted in the class of non-idling policies and the relative values of the holding cost rates did not matter in the analysis. Consequently, the results obtained in this section are novel for fully collaborative servers as well.

Taking into account Proposition 2, the optimality of idling policies is established in the following theorem.
\begin{theorem}
Assuming $h_1<h_2$, for each $x_1\geq 1$ there exists an integer $t(x_1)\geq 1$ such that $f(x_1,x_2)<0$ for $x_2\geq t(x_1)$.
\end{theorem}
\begin{IEEEproof}
It suffices to show that $f(x_1,x_2)$ is decreasing in $x_2$ for $x_2$ sufficiently large and $\lim_{x_2\rightarrow\infty}f(x_1,x_2)=-\infty$. The proof is by induction on $x_1$. For $x_2\geq 1$ we have
\begin{eqnarray*}
f(1,x_2+1)&=&V(1,x_2+1)-V(0,x_2+2)\\
&&\hspace{-2.8cm}\leq h_1+h_2(x_2+1)+W_{0,\nu_2+\mu_2}(1,x_2+1)-V(0,x_2+2)\\
&&\hspace{-2.8cm}=h_1-h_2+(\nu_2+\mu_2)V(1,x_2)+(1-\nu_2-\mu_2)V(1,x_2+1)\\
&&\hspace{-2.55cm}-(\nu_2+\mu_2)V(0,x_2+1)-(1-\nu_2-\mu_2)V(0,x_2+2)\\
&&\hspace{-2.8cm}=h_1-h_2+(\nu_2+\mu_2)f(1,x_2)+(1-\nu_2-\mu_2)f(1,x_2+1)\\
&&\hspace{-2.8cm}\Longrightarrow(\nu_2+\mu_2)\left[f(1,x_2+1)-f(1,x_2)\right]\leq h_1-h_2<0,
\end{eqnarray*}
which proves the result for $x_1=1$ and establishes the induction base. Assume that the result holds for some $x_1>1$, which implies that there exists $t(x_1)$
such that the optimal allocation for $x_2\geq \max\{t(x_1),2\}$ is $(0,\nu_2+\mu_2)$. Then, for $x_2\geq\max\{t(x_1)-2,1\}$ we can replicate the arguments used for $x_1=1$ to show that
\begin{displaymath}
(\nu_2+\mu_2)\left[f(x_1+1,x_2+1)-f(x_1+1,x_2)\right]\leq h_1-h_2<0,
\end{displaymath}
which completes the induction and the proof.
\end{IEEEproof}

With idling included there are three possible server allocations when there are jobs in both stations, so the optimal policy cannot be determined from the sign of a single decision function as was the case when $h_1\geq h_2$; if $f(x_1,x_2)<0$, the optimal policy idles the dedicated server of Station 1 and assigns the flexible server to Station 2, whereas if $f(x_1,x_2)\geq0$, the flexible server allocation is determined by one of the decision functions defined in (11)-(14). Moreover, it is not possible to use the techniques of the previous section, which were specific to the case of ${\cal A}(x_1,x_2)$ having two elements, to obtain a recursive expression for the decision function. For these reasons we were only able to characterize the optimal policy for two special cases of the general model: i) one job in Station 1, and ii) no dedicated server in Station 2.

In the following theorem we show that when there is one job in Station 1 the optimal policy is determined by two switching points.
\begin{theorem}
Assume $h_1<h_2$, $x_1=1$, and i) fully collaborative servers or ii) partially collaborative servers and $\nu_2\geq\mu_2$.
Then, there exist integers $t_2\geq t_1\geq1$ such that the optimal policy idles the dedicated server in Station 1 when $x_2\geq t_2$ and assigns the flexible server to Station 2 (resp. 1) when $x_2\geq t_1$ (resp. $x_2<t_1$).
\end{theorem}
\begin{IEEEproof}
The existence of $t_2$ follows from $f(1,x_2)$ being decreasing (see proof of Theorem 4). For $x_2<t_2$ non-idling policies are optimal so that the optimal allocation of the flexible server depends on the sign of $d(1,x_2)$ for fully collaborative servers and $\bar d(1,1)$, $\tilde d(1,x_2)$, $x_2>1$, for partially collaborative servers. Then, the existence of the lower switching point $t_1$ follows from $d(1,x_2)$ being decreasing (known from past work) and Lemma 4(i) for fully and partially collaborative servers, respectively.
\end{IEEEproof}

For partially collaborative servers we obtained numerical results indicating that condition $\nu_2\geq\mu_2$ is only needed in the proof.
We also believe that the theorem is valid for more than one job in Station 1, that is, for a fixed number of jobs in Station 1 the optimal policy is determined by two switching points $t_2(x_1)\geq t_1(x_1)\geq1$. Our conjecture was verified by extensive numerical experiments but we were able to prove it only when there is no dedicated server assigned to Station 2. The optimal policy for this case is given in the following theorem.
\begin{theorem}
Let $h_1<h_2$ and $\nu_2=0$. Then\\
i) The optimal policy assigns the flexible server to Station 2, that is, $t_1(x_1)=1$.\\
ii) For each $x_1\geq 1$, there exists an integer
$t_2(x_1)\geq 1$ such that the optimal policy idles the dedicated server of Station 1 when $x_2\geq t_2(x_1)$. Moreover, $t_2(x_1)$ in nondecreasing.
\end{theorem}

\section{Conclusion}
We characterized optimal server allocations for Markovian two-stage tandem queueing systems with dedicated servers in each stage and one flexible server. We considered server synergy models that included partial collaboration of servers working on the same job in addition to full collaboration that was the standard assumption in most of previous related work. We obtained novel results for any type of collaboration as well as extensions to the partial collaboration case of known results for fully collaborative servers. We noticed that the problem is more complex for partially collaborative servers and its technical analysis much more difficult. We also showed by examples that the partial collaboration assumption may alter significantly the structure of the optimal policy resulting in policies that do not possess intuition-based properties that have been shown to hold for fully collaborative servers.


%
\appendix

\section*{Proof of Lemma 4}
We first derive a recursive equation for $\tilde d(1,x_2)$, $x_2\geq0$. To do that we use optimality equations (3) and (5)-(8) and, where applicable, identities $\min(a,b)=a+(b-a)^-$ and $\min(a,b)=b-(b-a)^+$ for the first and second terms, respectively, of the differences appearing in the definition of $f(1,x_2)$ and $g(1,x_2)$. We get
\begin{eqnarray}
\tilde d(1,0)&=&\xi_1(h_1-h_2)+\xi_1(\nu_2+\xi_2)V(0,1)\nonumber\\
&&+(\nu_2+\mu_1+\mu_2+\xi_2)\tilde d(1,0),\\
\tilde d(1,1)&=&\xi_1(h_1-h_2)-\mu_2h_2+\nu_2\tilde d(1,0)+\mu_1\tilde d(1,1)\nonumber\\
&&+\nu_1\mu_2g(0,2)+\xi_1\xi_2f(1,1)+\mu_2^2g(1,1)\nonumber\\
&&+\xi_1\bar d(1,1)^-+\mu_2\bar d(1,1)^+,\\
\tilde d(1,x_2)&=&\xi_1(h_1-h_2)-\mu_2h_2+\nu_2\tilde d(1,x_2-1)\nonumber\\
&&\hspace{-1.5cm}+(\mu_1+\xi_2)\tilde d(1,x_2)+\nu_1\mu_2g(0,x_2+1)+\xi_1\tilde d(1,x_2)^-\nonumber\\
&&\hspace{-1.5cm}+\mu_2\tilde d(1,x_2)^++\mu_2\left[\bar d(1,1)^-{\bf 1} (x_2=2) \right.\nonumber\\
&&\hspace{-1.5cm}\left.+\tilde d(1,x_2-1)^-{\bf 1} (x_2>2)\right],~~x_2\geq 2.
\end{eqnarray}
We also note that $g(0,x_2)$ is negative and decreasing, a fact that follows easily from (5) and (6).

To prove $\tilde d(1,0)>\tilde d(1,1)$ we consider two cases for $\bar d(1,1)$.
When $\bar d(1,1)<0$ we get from (16)
\begin{eqnarray}
\tilde d(1,1)&=&\xi_1(h_1-h_2)-\mu_2h_2+\nu_2\tilde d(1,0)\nonumber\\
&&\hspace{-2cm}+(\mu_1+\mu_2)\tilde d(1,1)+\nu_1\mu_2g(0,2)+\xi_1(\xi_2-\mu_2)f(1,1)+\xi_1\bar d(1,1)\nonumber\\
&&\hspace{-2cm}<\xi_1(h_1-h_2)-\mu_2h_2+\nu_2\tilde d(1,0)+(\mu_1+\mu_2)\tilde d(1,1)
\end{eqnarray}
because $g(0,2)<0$, $\mu_2>\xi_2$, $f(1,1)>0$, and $\bar d(1,1)<0$. Then (15) and (18) yield
\begin{eqnarray*}
\left( 1-\mu_1-\mu_2\right)\left[\tilde d(1,0)-\tilde d(1,1)\right]&&\\
&&\hspace{-5cm}>\mu_2h_2+\xi_1(\nu_2+\xi_2)V(0,1)+\xi_2\tilde d(1,0)>0
\end{eqnarray*}
because $\tilde d(1,0)=\xi_1f(1,0)>0$.
When $\bar d(1,1)\geq0$, substituting $\bar d(1,1)^+=\bar d(1,1)=\xi_1f(1,1)+\xi_2g(1,1)$ in (16) we get
\begin{eqnarray*}
\tilde d(1,1)&=&\xi_1(h_1-h_2)-\mu_2h_2+\nu_2\tilde d(1,0)\\
&&+(\mu_1+\mu_2+\xi_2)\tilde d(1,1)+\nu_1\mu_2g(0,2),
\end{eqnarray*}
which combined with (15) gives
\begin{eqnarray*}
\left(1-\mu_1-\mu_2-\xi_2\right)\left[ \tilde d(1,0)-\tilde d(1,1)\right]&&\\
&&\hspace{-5.5cm}=\mu_2h_2+\xi_1(\nu_2+\xi_2)V(0,1)-\nu_1\mu_2g(0,2)>0.
\end{eqnarray*}
Next we show $\tilde d(1,1)>\tilde d(1,2)$. Using $\bar d(1,1)=\bar d(1,1)^-+\bar d(1,1)^+$ in (16) we get
\begin{eqnarray}
\tilde d(1,1)&=&\xi_1(h_1-h_2)-\mu_2h_2+\nu_2\tilde d(1,0)\nonumber\\
&&\hspace{-1cm}+(\mu_1+\xi_2)\tilde d(1,1)+\nu_1\mu_2g(0,2)\nonumber\\
&&\hspace{-1cm}+\mu_2\tilde d(1,1)+\xi_1\bar d(1,1)^--\mu_2\bar d(1,1)^-.
\end{eqnarray}
Assuming that $\tilde d(1,1)<0$ and taking into account that $\tilde d(1,1)<\bar d(1,1)$ and $\tilde d(1,1)^+=0$, we get from (19) and (17)
\begin{eqnarray*}
\tilde d(1,1)-\tilde d(1,2)&>&\nu_2\tilde d(1,0)+(\mu_2-\nu_2)\tilde d(1,1)\\
&&\hspace{-1.2in}+(\mu_1+\xi_2)\left[\tilde d(1,1)-\tilde d(1,2)\right]+\mu_1\nu_2\left[ g(0,2)-g(0,3)\right]\\
&&\hspace{-1.2in}+\xi_1\left[\tilde d(1,1)^--\tilde d(1,2)^-\right]+\mu_2\left[\tilde d(1,1)^+-\tilde d(1,2)^+\right]>0
\end{eqnarray*}
because of $\mu_2\leq\nu_2$, $g(0,x_2)$ being decreasing, and Lemma 2. When $\tilde d(1,1)\geq 0$, which implies that $\bar d(1,1)\geq 0$ as well, we get
\begin{eqnarray*}
\tilde d(1,1)-\tilde d(1,2)&=&\nu_2\left[\tilde d(1,0)-\tilde d(1,1)\right]\\
&&\hspace{-1.2in}+(\mu_1+\xi_2)\left[\tilde d(1,1)-\tilde d(1,2)\right]+\mu_1\nu_2\left[ g(0,2)-g(0,3)\right]\\
&&\hspace{-1.2in}+\mu_2\left[\tilde d(1,1)^+-\tilde d(1,2)^+\right]-\xi_1\tilde d(1,2)^->0,
\end{eqnarray*}
which follows from $\tilde d(1,0)>\tilde d(1,1)$, $g(0,x_2)$ being decreasing, and Lemma 2.
For $x_2\geq 2$, $\tilde d(1,x_2)-\tilde d(1,x_2+1)>0$ is proved by a straightforward induction on $x_2$ based on (17).

We now turn to the proof of part (ii). Because $\tilde d(1,x_2)$ is a decreasing sequence, its limit as $x_2\rightarrow\infty$ exists. Then, assuming that $\tilde L=\lim_{x_2\rightarrow\infty}\tilde d(1,x_2)$
is finite and taking limits in (17) we get
\begin{eqnarray*}
\tilde L&=&\xi_1(h_1-h_2)-\mu_2h_2+(\nu_2+\mu_1+\xi_2)\tilde L\\
&&\hspace{-1.5cm}+\nu_1\mu_2\lim_{x_2\rightarrow\infty}g(0,x_2)+\xi_1
\left(\tilde L-\tilde L^+\right)+\mu_2\left(\tilde L^++\tilde L^-\right)\\
&&\hspace{-1.5cm}\Rightarrow\nu_1\tilde L=\xi_1(h_1-h_2)-\mu_2h_2-\xi_1\tilde L^++\nu_1\mu_2\lim_{x_2\rightarrow\infty}g(0,x_2),
\end{eqnarray*}
which is a contradiction because $g(0,x_2)=-h_2x_2/(\nu_2+\mu_2)$.

\section*{Proof of Lemma 5}
The proof is by induction on $x_1$. For $x_1=1$ we derive the following recursive equation for $d(1,x_2)$.
\begin{eqnarray}
d(1,0)&=&\mu_1(h_1-h_2)+\mu_1(\nu_2+\xi_2)V(0,1)\nonumber\\
&&+(\nu_2+\mu_1+\mu_2+\xi_2)d(1,0),
\end{eqnarray}
and for $x_2\geq1$
\begin{eqnarray}
d(1,x_2)&=&\mu_1(h_1-h_2)-\mu_2h_2+\nu_2 d(1,x_2-1)\nonumber\\
&&\hspace{-1.8cm}+(\mu_1+\mu_2+\xi_2) d(1,x_2)+\mu_2(\nu_1+\xi_1-\mu_1)g(0,x_2+1)\nonumber\\
&&\hspace{-1.8cm}+(\mu_1-\mu_2)\left[\bar d(1,1)^-{\bf 1}(x_2=1)+\tilde d(1,x_2)^-{\bf 1}(x_2>1)\right]\nonumber\\
&&\hspace{-1.8cm}+\mu_2\left[\bar d(1,1)^-{\bf 1} (x_2=2) +\tilde d(1,x_2-1)^-{\bf 1} (x_2>2)\right].
\end{eqnarray}
Then $d(1,x_2)-d(1,x_2+1)>0$ can be proved by induction on $x_2$ based on (20) and (21), using the facts that $\nu_1+\xi_1> \mu_1\geq\mu_2$, $g(0,x_2)$ is negative and decreasing, $\bar d(1,1)>\tilde d(1,1)$, and $\tilde d(1,x_2)$ is decreasing.
For $x_1\geq 2$ we have
\begin{eqnarray}
d(x_1,0)&=&\mu_1(h_1-h_2)+\nu_1\mu_1f(x_1-1,1)\nonumber\\
&&\hspace{-1.9cm}-\mu_1(\nu_2+\xi_2)g(x_1-1,1)+(\nu_2+\mu_2+\xi_1+\xi_2)d(x_1,0)\nonumber\\
&&\hspace{-1.9cm}+\mu_1\left[\bar d(1,1)^+{\bf 1}(x_1=2)+\hat d(x_1-1,1)^+{\bf 1}(x_1>2)\right],\\
d(x_1,1)&=&\mu_1(h_1-h_2)-\mu_2h_2\nonumber\\
&&\hspace{-1.9cm}+\nu_1d(x_1-1,2)+\nu_2 d(x_1,0)+(\xi_1+\xi_2) d(x_1,1)\nonumber\\
&&\hspace{-1.9cm}+\mu_2(\mu_2-\xi_2)g(x_1,1)+\mu_1\hat d(x_1,1)^-+\mu_2\hat d(x_1,1)^+\nonumber\\
&&\hspace{-1.9cm}+\mu_1\left[\tilde d(1,2)^+{\bf 1}(x_1=2)+d(x_1-1,2)^+{\bf 1}(x_1>2)\right],\\
d(x_1,x_2)&=&\mu_1(h_1-h_2)-\mu_2h_2+\nu_1d(x_1-1,x_2+1)\nonumber\\
&&\hspace{-2.1cm}+\nu_2 d(x_1,x_2-1)+(\xi_1+\xi_2)d(x_1,x_2)+\mu_1 d(x_1,x_2)^-\nonumber\\
&&\hspace{-2.1cm}+\mu_2 d(x_1,x_2)^++\mu_1\tilde d(1,x_2+1)^+{\bf 1} (x_1=2)\nonumber\\
&&\hspace{-2.1cm}+\mu_1d(x_1-1,x_2+1)^+{\bf 1} (x_1>2)+\mu_2\hat d(x_1,1)^-{\bf 1} (x_2=2)\nonumber\\
&&\hspace{-2.1cm}+\mu_2 d(x_1,x_2-1)^-{\bf 1} (x_2>2),~~x_2\geq2.
\end{eqnarray}
To show $d(x_1,0)>d(x_1,1)>d(x_1,2)$ we first assume that $\hat d(x_1,1)<0$ which implies $d(x_1,1)<0$ as well. Then, $d(x_1,1)<d(x_1,0)$ because $d(x_1,0)=\mu_1f(x_1,0)>0$. Noting that $(\mu_2-\xi_2)g(x_1,1)=d(x_1,1)-
\hat d(x_1,1)$, $\hat d(x_1,1)>d(x_1,1)$, and $\hat d(x_1,1)^+=d(x_1,1)^+=0$, we get from (23) and (24)
\begin{eqnarray}
d(x_1,1)-d(x_1,2)&>&\nu_1\left[ d(x_1-1,2)-d(x_1-1,3)\right]\nonumber\\
&&\hspace{-1.2in}+\nu_2d(x_1,0)+(\mu_2-\nu_2)d(x_1,1)\nonumber\\
&&\hspace{-1.2in}+\mu_1\left[\tilde d(1,2)^+-\tilde d(1,3^+)\right] {\bf 1}(x_1=2)\nonumber\\
&&\hspace{-1.2in}+\mu_1\left[ d(x_1-1,2)^+-d(x_1-1,3)^+\right]{\bf 1}(x_1>2)\nonumber\\
&&\hspace{-1.2in}+(\xi_1+\xi_2)\left[ d(x_1,1)-d(x_1,2)\right]+\mu_1\left[ d(x_1,1)^- -d(x_1,2)^-\right]\nonumber\\
&&\hspace{-1.2in}+\mu_2\left[ d(x_1,1)^+ -d(x_1,2)^+\right].
\end{eqnarray}
Because $d(x_1,0)>0$, $\mu_2\leq\nu_2$, $d(x_1,1)<0$, and $\tilde d(1,x_2)$ is decreasing, we get $d(x_1,1)>d(x_1,2)$ by applying the induction hypothesis and Lemma 2. When $\hat d(x_1,1)\geq 0$, we have in (23) $\mu_2(\mu_2-\xi_2)g(x_1,1)+
\mu_2\hat d(x_1,1)^+=\mu_2d(x_1,1)$. Therefore, taking also into account that $\mu_1f(x_1-1,1)>d(x_1-1,1)$ and $g(x_1-1,1)<0$, we obtain from (22) and (23)
\begin{eqnarray*}
(1-\xi_1-\xi_2-\mu_2)\left[ d(x_1,0)-d(x_1,1)\right]&&\\
&&\hspace{-2.1in}>\mu_2h_2+\nu_1\left[ d(x_1-1,1)-d(x_1-1,2)\right]\\
&&\hspace{-2in}+\mu_1\left[\bar d(1,1)^+-\tilde d(1,2)^+\right] {\bf 1}(x_1=2)\\
&&\hspace{-2in}+\mu_1\left[ \hat d(x_1-1,1)^+ -d(x_1-1,2)^+\right]{\bf 1}(x_1>2).
\end{eqnarray*}
The righthand side of the equation above is positive by the induction hypothesis, $\bar d(1,1)>\tilde d(1,1)>\tilde d(1,2)$, and $\hat d(x_1-1,1)>d(x_1-1,1)$ for $x_1>2$. When $d(x_1,1)<0$, (23) and (24) yield (25) without the second to last term, so $d(x_1,1)>d(x_1,2)$ is proved similarly. When $d(x_1,1)\geq 0$, in which case $d(x_1,1)=d(x_1,1)^+$, we get
\begin{eqnarray*}
d(x_1,1)-d(x_1,2)&>&\nu_1\left[ d(x_1-1,2)-d(x_1-1,3)\right]\\
&&\hspace{-1.3in}+\nu_2\left[ d(x_1,0)-d(x_1,1)\right]+\mu_1\left[\tilde d(1,2)^+-\tilde d(1,3)^+\right] {\bf 1}(x_1=2)\\
&&\hspace{-1.3in}+\mu_1\left[ d(x_1-1,2)^+-d(x_1-1,3)^+\right]{\bf 1}(x_1>2)\\
&&\hspace{-1.3in}+(\xi_1+\xi_2)\left[ d(x_1,1)-d(x_1,2)\right]+\mu_2\left[ d(x_1,1)^+ -d(x_1,2)^+\right],
\end{eqnarray*}
and $d(x_1,1)>d(x_1,2)$ follows from $d(x_1,0)>d(x_1,1)$, $\tilde d(1,x_2)$ being decreasing, the induction hypothesis, and Lemma 2.
For $x_2\geq 2$, $d(x_1,x_2)-d(x_1,x_2+1)>0$ follows from (24) by applying induction on $x_2$ and using the induction hypothesis for $x_1$.

For part (ii) we let $L(x_1)=\lim_{x_2\rightarrow\infty}d(x_1,x_2)$ and use induction on $x_1$. From (21) we have
\begin{eqnarray*}
d(1,x_2)&\leq&\mu_1(h_1-h_2)-\mu_2h_2+\nu_2 d(1,x_2-1)\\
&&\hspace{-2cm}+(\mu_1+\mu_2+\xi_2) d(1,x_2)+\mu_2(\nu_1+\xi_1-\mu_1)g(0,x_2+1).
\end{eqnarray*}
Assuming $L(1)$ is finite and taking limits on both sides we get
\begin{eqnarray*}
(\nu_1+\xi_1)L(1)&\leq&\mu_1(h_1-h_2)-\mu_2h_2\\
&&\hspace{-1cm}+\mu_2(\nu_1+\xi_1-\mu_1)\lim_{x_2\rightarrow\infty}g(0,x_2)=-\infty,
\end{eqnarray*}
clearly a contradiction. Assuming that $L(x_1-1)=-\infty$ (induction hypothesis) and taking also into account that $\tilde L=-\infty$, we get from (24) for $x_2$ sufficiently large
\begin{eqnarray*}
d(x_1,x_2)&=&\mu_1(h_1-h_2)-\mu_2h_2+\nu_1d(x_1-1,x_2+1)\\
&&\hspace{-2cm}+\nu_2 d(x_1,x_2-1)+(\xi_1+\xi_2)d(x_1,x_2)+\mu_1 d(x_1,x_2)^-\\
&&\hspace{-2cm} +\mu_2 d(x_1,x_2)^++\mu_2 d(x_1,x_2-1)^-.
\end{eqnarray*}
Assuming $L(x_1)$ is finite and taking limits on both sides we get
\begin{eqnarray*}
\nu_1L(x_1)&=&\mu_1(h_1-h_2)-\mu_2h_2\\
&&\hspace{-1cm}-\mu_1L(x_1)^++\nu_1L(x_1-1)=-\infty,
\end{eqnarray*}
which is a contradiction, completing the induction and the proof.

\section*{Proof of Lemma 6}
The proof of part (i) is by induction on $x_2$. Note that $d(2,1)\geq0$ implies $\hat d(2,1)\geq 0$, and for $x_2\geq2$, $d(2,x_2)\geq0$ implies $d(2,x_2-1)\geq 0$ by Lemma 5(i). Taking the above into account and after some straightforward algebra we get from (23), (24), and (21)
\begin{eqnarray*}
d(2,x_2)&\geq&\mu_1(h_1-h_2)-\mu_2h_2+\nu_1d(1,x_2+1)\\
&&\hspace{-1cm}+\nu_2d(2,x_2-1)+(\xi_1+\xi_2+\mu_2)d(2,x_2), \\
d(1,x_2+1)&\leq&\mu_1(h_1-h_2)-\mu_2h_2+\nu_2d(1,x_2)\\
&&+(\mu_1+\mu_2+\xi_2)d(1,x_2+1),
\end{eqnarray*}
from which we obtain
\begin{eqnarray}
(\nu_1+\nu_2+\mu_1)\left[d(2,x_2)-d(1,x_2+1)\right]&&\nonumber\\
&&\hspace{-2.5in}\geq(\nu_1+\xi_1-\mu_1) d(1,x_2+1)+\nu_2\left[ d(2,x_2-1)-d(1,x_2)\right].
\end{eqnarray}
We assume that $d(1,x_2+1)\geq0$ because otherwise there is nothing to prove. Then, $d(2,1)\geq d(1,2)$ follows from $d(2,0)>d(1,0)>d(1,1)$ and (26) for $x_2=1$, establishing the induction base. For $x_2\geq2$, because $d(2,x_2-1)\geq 0$, the induction hypothesis implies that $d(2,x_2-1)-d(1,x_2)\geq0$, so we get $d(2,x_2)-d(1,x_2+1)\geq0$ from (26).

Before proceeding to parts (ii) and (iii), we use the optimality equations to get for $x_1\geq2$
\begin{eqnarray}
\hat d(x_1,1)&=&C(x_1)+(\xi_1+\mu_2)\hat d(x_1,1)\nonumber\\
&&+\mu_1\hat d(x_1,1)^-+\xi_2\hat d(x_1,1)^+,
\end{eqnarray}
where
\begin{eqnarray}
C(x_1)&=&\mu_1(h_1-h_2)-\xi_2h_2\nonumber\\
&&\hspace{-1.7cm}+\nu_1\left[ \mu_1f(x_1-1,2)+\xi_2g(x_1-1,2) \right]\nonumber\\
&&\hspace{-1.7cm}+\nu_2d(x_1,0)-\mu_1(\mu_2-\xi_2)g(x_1-1,2)\nonumber\\
&&\hspace{-1.7cm}+\mu_1\left[\tilde d(1,2)^+{\bf 1}(x_1=2)+d(x_1-1,2)^+{\bf 1}(x_1>2)\right]\nonumber\\
&&\hspace{-1.95cm}>\mu_1(h_1-h_2)-\xi_2h_2+\nu_2d(x_1,0)+(\nu_1+\mu_1)\nonumber\\
&&\hspace{-1.7cm}\times\left[ \tilde d(1,2)^+{\bf 1}(x_1=2)+d(x_1-1,2)^+{\bf 1}(x_1>2)\right],
\end{eqnarray}
with the inequality following from $\mu_i>\xi_i$, $i=1,2$.

The proof of part (ii) is by induction on $x_2$. We have $\bar d(1,1)>\tilde d(1,1)>\tilde d(1,2)$ with the second inequality following from Lemma 4(i). Therefore, assuming that $\tilde d(1,2)\geq0$, we get from (17)
\begin{eqnarray}
(\nu_1+\nu_2+\xi_1)\tilde d(1,2)&=&\xi_1(h_1-h_2)-\mu_2h_2\nonumber\\
&&\hspace{-1cm}+\nu_2\tilde d(1,1)+\nu_1\mu_2g(0,3)\geq0.
\end{eqnarray}
Because $d(2,0)>d(1,1)>\tilde d(1,1)$, we get that $C(2)$ is larger than the righthand side of (29). Therefore, $C(2)>0$ and we get $\hat d(2,1)>0$ from (27) and Lemma 2. Substituting in (23) we get
\begin{eqnarray*}
d(2,1)&=&\mu_1(h_1-h_2)-\mu_2h_2+\nu_1d(1,2)+\nu_2d(2,0)\\
&&+\mu_1\tilde d(1,2)+(\xi_1+\xi_2+\mu_2)d(2,1),
\end{eqnarray*}
and in combination with (29)
\begin{eqnarray*}
(\nu_1+\nu_2+\mu_1)\left[d(2,1)-\tilde d(1,2)\right]&>&(\mu_1-\xi_1)(h_1-h_2)\\
&&\hspace{-2in}+\nu_1d(1,2)+\xi_1\tilde d(1,2)+\nu_2\left[ d(2,0)-\tilde d(1,1)\right],
\end{eqnarray*}
which is positive because $d(1,2)>\tilde d(1,2)\geq0$ and $d(2,0)>\tilde d(1,1)$, thus establishing the induction base. For $x_2>2$, $\tilde d(1,x_2)\geq0$ implies $\tilde d(1,x_2-1)\geq0$ by Lemma 4(i), which by the induction hypothesis yields $d(2,x_2-2)\geq0$. Taking into account all of the above we get from (24) and (17)
\begin{eqnarray*}
d(2,x_2-1)-\tilde d(1,x_2)&=&(\mu_1-\xi_1)(h_1-h_2)\\
&&\hspace{-1.2in}+\nu_1d(1,x_2)+\xi_1\tilde d(1,x_2)-\nu_1\mu_2g(0,x_2+1)\\
&&\hspace{-1.2in}+\nu_2\left[ d(2,x_2-2)-\tilde d(1,x_2-1)\right]\\
&&\hspace{-1.2in}+(\xi_1+\xi_2)\left[ d(2,x_2-1)-\tilde d(1,x_2)\right]\\
&&\hspace{-1.2in}+\mu_1\left[ d(2,x_2-1)^--\tilde d(1,x_2)^-\right]\\
&&\hspace{-1.2in}+\mu_2\left[ d(2,x_2-1)^+-\tilde d(1,x_2)^+\right].
\end{eqnarray*}
Noting that $d(1,x_2)>\tilde d(1,x_2)\geq0$ and applying the induction hypothesis to the term multiplying $\nu_2$, we get from Lemma 2 that $d(2,x_2-1)-\tilde d(1,x_2)\geq0$.

The proof of part (iii) is by nested induction on $x_1,x_2$. For some $x_1\geq 2$, assume that $d(x_1,2)\geq0$. Then, $d(x_1,1)>0$ by Lemma 5(i), implying $\hat d(x_1,1)>0$ as well. Therefore, we get from (24)
\begin{eqnarray}
(\nu_1+\nu_2+\mu_1)d(x_1,2)&=&\mu_1(h_1-h_2)-\mu_2h_2\nonumber\\
&&\hspace{-3cm}+\nu_1d(x_1-1,3)+\nu_2d(x_1,1)+\mu_1\left[\tilde d(1,3)^+{\bf 1}(x_1=2)\right.\nonumber\\
&&\hspace{-3cm}\left.+d(x_1-1,3)^+{\bf 1}(x_1>2)\right]\geq0.
\end{eqnarray}
Assuming that the lemma holds for less than $x_1$ jobs in Station 1 (induction hypothesis with respect to $x_1$), we get $d(x_1,2)\geq d(x_1-1,3)$ if $x_1>2$. If $x_1=2$ we have $d(2,2)\geq d(1,3)>\tilde d(1,3)$, where the first inequality is due to part (i). Moreover, $d(x_1+1,0)>d(x_1,0)>d(x_1,1)$, so (28) and (30) yield $C(x_1+1)>0$, and $\hat d(x_1+1,1)>0$ follows from (27) and Lemma 2. Substituting in (23) we get
\begin{eqnarray}
d(x_1+1,1)&=&\mu_1(h_1-h_2)-\mu_2h_2+\nu_1d(x_1,2)\nonumber\\
&&\hspace{-2.8cm}+\nu_2d(x_1+1,0)+\mu_1d(x_1,2)+(\xi_1+\xi_2+\mu_2)d(x_1+1,1).
\end{eqnarray}
Using part (i) for $x_1=2$ and the induction hypothesis for $x_1>2$ we get from (30) and (31) that $d(x_1,2)\leq d(x_1+1,1)$, which establishes the base for the induction with respect to $x_2$. The induction is completed by using (24) to get an expression for $d(x_1+1,x_2-1)-d(x_1,x_2)$, $x_1\geq2$, $x_2\geq3$, and then reasoning as in the case $x_2=2$.

\section*{Proof of Theorem 6}
When $f(x_1,x_2)<0$, Propositions 1 and 2 imply that the dedicated server in Station 1 should be idled and the flexible server should be assigned to Station 2. Therefore, to prove the first part of the theorem we only need to show that decision functions $\tilde d(1,x_2)$ and $d(x_1,x_2)$ are negative for $x_1\geq2$, $x_2\geq1$ such that $f(x_1,x_2)\geq 0$.

Because $f(1,x_2)$ is a decreasing sequence (see proof of Theorem 4), there exists $x_2^*$ such that $f(1,x_2)\geq 0$ for $x_2\leq x_2^*$. If $x_2^*=0$, there is nothing to prove. Otherwise, we use optimality equations (8) and (10) with $\nu_2=\xi_2=0$ to get
\begin{eqnarray*}
\tilde d(1,x_2)&=&\xi_1(h_1-h_2)-\mu_2h_2+\mu_1\tilde d(1,x_2)\\
&&\hspace{-1cm}+\nu_1\mu_2g(0,x_2+1)+\xi_1\tilde d(1,x_2)^-+\mu_2\tilde d(1,x_2)^+\\
&&\hspace{-1cm}+\mu_2\tilde d(1,x_2-1){\bf 1}(x_2>1),~~1\leq x_2\leq x_2^*,
\end{eqnarray*}
and $\tilde d(1,x_2)<0$ can be proved by a straightforward induction on $x_2$. Therefore, the optimal allocation in state $(1,x_2)$, $x_2\leq x_2^*$, is $(\nu_1,\mu_2)$, and optimality equations (1) and (2) give
\begin{eqnarray}
d(1,1)&=&\mu_1(h_1-h_2)-\mu_2h_2+(\mu_1+\xi_1)d(1,1)\nonumber\\
&&+\nu_1\mu_2g(0,2)+\mu_2(\mu_1-\xi_1)f(1,0),\\
d(1,x_2)&=&\mu_1(h_1-h_2)-\mu_2h_2+(\mu_1+\xi_1)d(1,x_2)\nonumber\\
&&\hspace{-1cm}+\nu_1\mu_2g(0,x_2+1)+\mu_2d(1,x_2-1),~~x_2\geq2.
\end{eqnarray}
>From (3) we have $f(1,0)=h_1/(\nu_1+\xi_1)$. Because $h_1<h_2$ and $\nu_1+\xi_1>\mu_1$ it is easily seen that $(\mu_1-\xi_1)f(1,0)-h_2<0$, and $d(1,1)<0$ follows from (32). For $x_2\geq 2$, $d(1,x_2)<0$ follows directly by applying induction in (33).

Next we show $d(x_1,x_2)<0$ for $x_1\geq 2$ by nested induction on $x_1,x_2$. Assume that $d(x_1-1,x_2)<0$ for $x_2\geq 1$ and $d(x_1,x_2-1)<0$ if $x_2>1$ (induction hypothesis). Note that for states $(y_1,y_2)$ with $d(y_1,y_2)<0$ the optimal allocation is either $(\nu_1,\mu_2)$ or $(0,\mu_2)$, resulting in the following optimality equation.
\begin{eqnarray}
V(y_1,y_2)&=&h_1y_1+h_2y_2+\mu_2V(y_1,y_2-1)\nonumber\\
&&\hspace{-1cm}+\min\{\nu_1V(y_1-1,y_2+1)+(\mu_1+\xi_1)V(y_1,y_2),\nonumber\\
&&\hspace{-1cm}(\nu_1+\mu_1+\xi_1)V(y_1,y_2)\}.
\end{eqnarray}
Assuming $f(x_1,x_2)\geq 0$, we use (10) for $V(x_1,x_2)$ and (34) for $V(x_1-1,x_2+1)$ and $V(x_1,x_2-1)$ to get an expression for $d(x_1,x_2)$. Noting that the difference of the two terms in braces in (34) is equal to $\nu_1f(y_1,y_2)$, we get for $x_2\geq1$
\begin{eqnarray}
d(x_1,x_2)&=&\mu_1(h_1-h_2)-\mu_2h_2\nonumber\\
&&\hspace{-2.5cm}+\nu_1\mu_1f(x_1-1,x_2+1)^++\nu_1\mu_2g(x_1-1,x_2+1)\nonumber\\
&&\hspace{-2.5cm}+\xi_1d(x_1,x_2)+\mu_1 d(x_1,x_2)^- +\mu_2 d(x_1,x_2)^+\nonumber\\
&&\hspace{-2.5cm}+\mu_2\left[d(x_1,x_2-1)+\nu_1f(x_1,x_2-1)^-\right]{\bf 1} (x_2>1).
\end{eqnarray}
Because $\mu_1f(x_1-1,x_2+1)^++\mu_2g(x_1-1,x_2+1)= \max\{d(x_1-1,x_2+1),\mu_2g(x_1-1,x_2+1)\}<0$, we obtain $d(x_1,x_2)<0$ by applying the induction hypothesis and Lemma 2 in (35).

For part (ii) it suffices to show that $f(x_1,x_2)$ is increasing in $x_1$ and decreasing in $x_2$. Taking into account that the optimal policy assigns the flexible server to the downstream station, we use (34) to derive the following recursive equation for $f(x_1,x_2)$, $x_1\geq1$, $x_2\geq1$.
\begin{eqnarray}
f(x_1,x_2)&=&h_1-h_2+(\mu_1+\xi_1)f(x_1,x_2)\nonumber\\
&&+\mu_2f(x_1,x_2-1)+\nu_1f(x_1,x_2)^-\nonumber\\
&&+\nu_1f(x_1-1,x_2+1)^+{\bf 1}(x_1>1).
\end{eqnarray}
The proof of the first monotonicity property, $f(x_1+1,x_2)>f(x_1,x_2)$, $x_1\geq1$, is by a straightforward induction on $x_1,x_2$ based on (36), with the induction base for each $x_1\geq1$ established by the fact that $f(x_1+1,0)>f(x_1,0)$, which follows from (3) and (4). The second monotonicity property has already been proved for $x_1=1$ (see proof of Theorem 4). For $x_1\geq2$ we can prove that $f(x_1,x_2)$ is decreasing in $x_2$ by similar induction arguments provided that we can also show that $f(x_1,1)<f(x_1,0)$ to establish the induction base for each $x_1$. For that purpose we use a sample path argument. Let $P1$ and $P2$ be the processes that start in states $(x_1,1)$ and $(x_1-1,2)$, respectively, and assume that the optimal policy, say $\pi$, is applied to $P2$. As for $P1$, we apply a policy $\bar\pi$ that imitates $\pi$ until the first time that Station 2 is empty under $P1$ and has one job under $P2$, and is optimal afterwards. Let $\tau$ be that time and $y_1$ be the number of jobs in Station 1 under $P1$ at time $\tau$. The two policies have a holding cost rate difference of $h_1-h_2$ until time $\tau$ and are optimal afterwards. Therefore, because $\bar\pi$ is not necessarily optimal we have
\begin{eqnarray}
V(x_1,1)-V(x_1-1,2)&\leq& (h_1-h_2)E(\tau)\nonumber\\
&&\hspace{-4cm}+E\left[V(y_1,0)-V(y_1-1,1)\right]<E\left[f(y_1,0)\right],
\end{eqnarray}
because $h_1<h_2$. Along every sample path we have $y_1\leq x_1$, so $f(y_1,0)\leq f(x_1,0)$, which combined with (37) yields $f(x_1,1)<f(x_1,0)$.





\ifCLASSOPTIONcaptionsoff
  \newpage
\fi

\end{document}